\documentclass[%
reprint,
amsmath,amssymb,
aps,
]{revtex4-2}

\usepackage{graphicx}% Include figure files
\usepackage{dcolumn}% Align table columns on decimal point
\usepackage{bm}% bold math
\usepackage[colorlinks=true,citecolor=blue,linkcolor=blue,urlcolor=blue]{hyperref}% add hypertext capabilities
\usepackage{tikz}
\usepackage{xcolor}
\usepackage{soul}

\newcommand{\orcidicon}[1]{\href{https://orcid.org/#1}{\includegraphics[height=\fontcharht\font`\B]{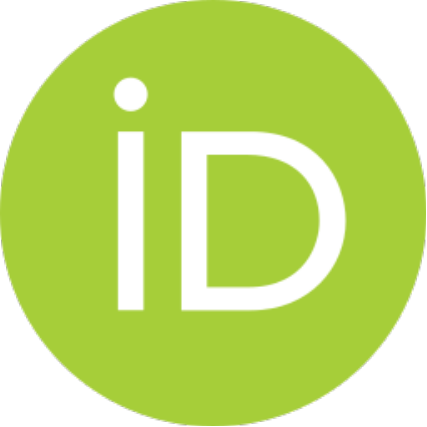}}}

\usepackage{mathrsfs}
\usepackage{pstricks}
\usepackage{color}
\usepackage{slashed}
\usepackage{epsfig}
\usepackage{amsfonts}
\def\beq{\begin{equation}}
\def\eeq{\end{equation}}
\def\bea{\begin{eqnarray}}
\def\eea{\end{eqnarray}}

\def\x{\mathbf{x}}

\def\q_perp{\mathbf{q}_{\perp}}
\begin{document}

\title{The Casimir-Lifshitz formula for rectangular dielectric waveguide}
\author{E. Arias\,\orcidicon{0000-0001-5064-8401}}
\email{email address: earias@iprj.uerj.br}
\affiliation{Universidade do Estado do Rio de Janeiro,  28625-570 Nova Friburgo, RJ, Brazil}

\author{G. O. Heymans\,\orcidicon{0000-0002-1650-4903 }}
\email{email address: olegario@cbpf.br}
\affiliation{Centro Brasileiro de Pesquisas F\'{\i}sicas, 22290-180 Rio de Janeiro, RJ, Brazil}

\author{N. F.~Svaiter\,\orcidicon{0000-0001-8830-6925}}
\email{email address: nfuxsvai@cbpf.br}
\affiliation{Centro Brasileiro de Pesquisas F\'{\i}sicas, 22290-180 Rio de Janeiro, RJ, Brazil}

%%%%%%%%%%%%%%%%%%%%%%%%%%%%%%%%%%%%%%%%

\begin{abstract}

%In this work we investigate the vacuum energy of the electromagnetic field inside a rectangular waveguide. We consider a $(3+1)$-dimensional spacetime and analyze the electromagnetic field in a waveguide compose by two different dielectric materials. We use the surface mode technique established in the literature to obtain a generalized Lifshitz formula for this geometry. We found the general expression for the Casimir energy and pressure and in the limit case we recover the usual formula for the perfect reflecting boundaries.

We analyze the Casimir-Lifshitz effect associated with the electromagnetic field in the presence of a rectangular waveguide consisting of two distinct dielectric materials in a $(3+1)$-dimensional spacetime. We employ the surface mode technique to derive a generalized Lifshitz formula for this specific geometry. Our formulation accounts for the unique dielectric properties of the materials composing the waveguide, leading to a precise calculation of the Casimir-Lifshitz energy. In the asymptotic limit, our results recover the classical expressions for perfect reflecting boundaries. 
%\st{The behavior of the force on the rectangular cavity walls admit us to identify the system as a Casimir-Lifshitz anharmonic oscillator, that may vibrate due only the to vacuum stress if we allow two parallel non-fixed walls}.
This work extends the applicability of the Lifshitz formula to more complex systems and provides valuable insights into the influence of dielectric materials on the electromagnetic Casimir effect.

\end{abstract}

%%%%%%%%%%%%%%%%%%%%%%%%%%%%%%%%%%%%%%%%


\maketitle

%%%%%%%%%%%%%%%%%%%%%%%%%%%%%%%%%%%%%%%%

\section{Introduction}\label{intro}

The Casimir effect is a remarkable consequence of the  
quantization of the electromagnetic field and the distortion of vacuum modes by the presence of classical boundaries. This effect demonstrates the influence of vacuum modes of quantum fields in macroscopic phenomena. Since the seminal work of Casimir \cite{Casimir:1948dh}, this effect has attracted great research interest  due to its wide connections with different areas of theoretical explorations and practical applications  \cite{Ambjorn:1981xw,Plunien:1986ca,Milonni, Bordag:2001qi,Lamoreaux_2005, bordag_book,  Klimchitskaya:2009cw}. From the theoretical perspective, one can mention the importance of zero-point energy in cosmology with 
standard model extension perspectives \cite{weinberg_cosmo_const}. Also, from the point of view of technological applications, the Casimir effect could bring us limits for  classical boundaries interference 
%and local properties of 
over quantum  effects in nanotechnology devices \cite{Woods:2015pla, nanophotonics2020}.
%The nanotechnology frontier that in recent years have been attended need the focus over the interference of classic frontier over quantum effects in real materials.
The experimental verification of Casimir prediction have been established along the years by numerous efforts \cite{Sparnaay:1958wg, Mohideen:1998iz,  Babb:2000yi, chan2001quantum}. These works confirm  the existence of forces, induced by vacuum fluctuations, between macroscopic objects. In this way, we mention, for example, the works of
Lamoreaux  where a micromechanical torsional oscillator have been employed to measure the Casimir pressure with high precision \cite{lamoreaux97}. 
%For other side, Mohideen \cite{Mohideen:1998iz} use an atomic force microscope and. 
%For exhaustive review on Casimir effect.
An interesting discussion of MEMS (MicrosElectroMechanical System) using vacuum force was presented by Serry et al. in Ref. \cite{ACOmaclay1995}. These authors consider an anharmonic Casimir oscillator constructed with a spring attached to one of the  plates of the Casimir slab configuration. 
%\st{Here, in the present work discussing a rectangular waveguide with different dielectric materials from inside and outside, we show that it is possible to construct the anharmonic Casimir-Lifshitz oscillator.
%We show that the finite conductivity corrections of the material can induce an effective potential well in the Casimir energy.
%{\color{red}and consequently the waveguide walls can exhibit natural vibrations due only to the vacuum stresses, if we allow two parallel non-fixed walls.}
%Near the zero-point energy local minimum, oscillatory systems may be constructed. One can discuss further the case where rigid plates are allowed to move. We are interpreting this as the Casimir-Lifshitz anharmonic oscillator.}

The Casimir effect in real materials is an important topic of modern research. In this way, the work of Lifshitz is a cornerstone that boosted the analysis of Casimir effect in real material media \cite{Lifshitz:1956zz}. The basic idea is to model vacuum quantum fluctuations by a stochastic fluctuating electromagnetic field and explore the fluctuation-dissipation theorem to infer the frequency dependent Casimir energy. The Lifshitz general result 
give us a formula for the Casimir energy density and pressure in a plane geometry fulfill with different dielectric media.
%dielectric model used in this article shows  a general . 
Other approaches have enriched the discussion and reinforce the result obtained by Lifshitz. For instance, by using Green's function approach, Dzyaloshinskii et al. obtain the same result for the Casimir effect in dielectrics \cite{DZYALOSHINSKII}. 
In the framework of spectral theory, it is possible to obtain the correction to Casimir energy due to finite conductivity effects, by using approximate functional equations. Recently it was obtained 
the Casimir
energy for a massless scalar field in the presence of a rectangular box  with non-ideal boundary conditions using an approximate functional equation  of the Epstein zeta-function \cite{nami2023}.
This derivation was possible using an analytic regularization procedure. For a discussion of the relation between an analytic regularization procedure and the cut-off method see \cite{Svaiter:1991je,nami1992,nami1993}.
%For other side, Barash and Ginzburg [] propose a method based on . 
%In order to review different approaches see Ref. .
We also mention 
the approach proposed by Van Kampen et al., 
%In that work the boundary conditions impose by the classical background geometry of the classical frontiers and its connection with the zero-point energy is made evident. 
where the physical effect of dielectric is carried on by the so call surface stationary modes of the electromagnetic field \cite{VANKAMPEN1968307}. 
%This technique is supported by further analysis [].
%Mostepanenko et al. \cite{mostepanenko2023} have discuss recently about a discrepancy between the real material models widely used in the literature. To be more specific i
 The discrepancy between the dissipationless plasma model of dielectric materials and the dissipative Drude model 
 was also discussed in the literature, e.g. Ref. ~\cite{mostepanenko2023}. 
%This author argue that a low frequency response of metals give rise to a ``Casimir puzzle" that confronts both models.
Although finite-sized cavity with perfect conductors have been discussed before, see Ref. \cite{Ambjorn:1981xw}, the inclusion of the dielectric lead to highly non-trivial problems. The case of a dielectric cylinder was discussed in Ref. \cite{Cavero-Pelaez:2004ydl}. However, the rectangular waveguide was not discussed, we believe, because the conditions to make the boundary conditions consistent alongside to the corner is problematic. Our approach to such problems will be discussed in this manuscript.

In the present work we 
%attempt to follow this line of research for 
study the Casimir effect in a rectangular waveguide that is filled with different dielectric materials using the Van Kampen method.
%\cite{VANKAMPEN1968307} different geometry. We investigate the Casimir effect in a 
%This general model lead us general formulas for Casimir energy and pressure that could be explore to distinguish the plasma and Drude model in the case of a real waveguide.
%Previous studies have focus on the influence of the geometry of the system over the Casimir effect. The work of Boyer and others \cite{Boyer:1968uf} on the spherical symmetry situation for Casimir effect, for example, rule out the possibility of a classical model for the electron as an spherical shell, since the sign of the force is not attractive in this geometry, but repulsive. A cylindrical geometry was first study for the electromagnetic field by Balian \cite{BALIAN1978165} and Milton \cite{milton_cylinder}. The case of a rectangular cavity of different length was study by Lucosz \cite{LUKOSZ1971109} and others indicates the sensibility of the attractive-repulsive nature of Casimir effect dependence on the geometry; see also Ref. \cite{Caruso:1990cgh}. It was shown that even for the same geometry (rectangular cavity)  an attractive force could become a repulsive force if the aspect ratio of the cavity lengths are modify.
The key point is the sensibility
of the attractive-repulsive nature of Casimir effect depending on the geometry or the dielectric characteristic of the material \cite{Boyer:1968uf, BALIAN1978165, milton_cylinder,LUKOSZ1971109, Caruso:1990cgh}. 
%Here we need to mention the exhaustive work of Wolfram and Ambjorn \cite{AMBJORN19831} where a rectangular symmetry is study in general dimensions ground and analytic techniques, allowing the possibility of arbitrary open or closed dimensions. 
%We need to mention, however, in its great majority, the studies for non-planar configurations have been performed by assuming ideal reflecting material, leaving a gap for study the geometry dependence in real material media. In this way, until the present, we will elaborate over the Casimir effect on a rectangular semi-open waveguide cavity for the electromagnetic field. A similar study for the ideal material was elaborate in Ref. \cite{Valuyan_2008}. 
%Here we use the V. Kampen methodology and look for a general Lifshitz result in a rectangular waveguide.
Our formulation accounts for the unique dielectric properties of the materials composing the waveguide, leading to a precise calculation of the Casimir-Lifshitz energy. 
%\st{The behavior of the force on the walls of the rectangular cavity allow us to identify the system as a Casimir-Lifshitz anharmonic oscillator, where the parallel walls can vibrate due only to vacuum stresses, if we allow two parallel non-fixed walls.}
One should notice that in the critical Casimir effect \cite{Dantchev:2022hvy}, in the presence of disorder, the sign of the Casimir force between boundaries depends in a non-trivial way on the strength of the non-thermal control parameter \cite{Rodriguez-Camargo:2022wyz,Heymans:2024dzq}.

%We like to point out that the correction to the Casimir force for imperfect boundaries conditions was recently derived using an approximate functional equation of the Riemann and Epstein zeta functions \cite{Arias:2023vfr}. 

The presentation of the work is as follows. In Section \ref{sec:perfect casimir} we discuss the Casimir energy in a perfect conducting waveguide. Section \ref{sec:stattionary modes} presents the stationary electromagnetic modes in dielectric waveguide, %are analyzed. There we used the Maxwell equations in order to obtain 
 and the surface modes frequencies are presented. %Then, i
We obtain the zero-point energy in the dielectric waveguide and the Casimir energy and force in Section \ref{sec:casimir energy}. %that  will be used to regularized to obtain finally 
%we also elaborate over some limits of our principal result and show the consistency with the previous results in the literature. 
Conclusion are presented in Section \ref{sec:conclusion}.

\section{Casimir energy in a perfect conducting waveguide}
\label{sec:perfect casimir}
In this section, we present the Casimir energy of a rectangular waveguide with perfect conductivity walls by using an analytic regularization procedure. For convenience, we consider a $(d+1)$ dimensional spacetime.
%Which means that we will obtain the expectation value of the Hamiltonian of the electromagnetic field satisfying Dirichlet boundary conditions in the two compacted dimensions. This is equivalent to say that we wish to 
We regularize the spectrum of the Hamiltonian in the spatial domain $\mathbb{R}^{d-2}\times[0,a]\times[0,b]$ \cite{Svaiter:1991je, Valuyan_2008}.
%. Here we present the calculations of References 

The quantization of the electromagnetic field can be performed using the vector potential. In this case, 
 %\begin{equation}
  %   L = \frac{1}{2} \left(\frac{\mathbf{E}^2}{c^2} - \mathbf{B}^2\right).
 %\end{equation}
%one can fix the gauge and easily obtain a equivalent expression using the vector potential. In such a situation, problems with the quantization can arise from the fact that 
the scalar and longitudinal modes are presented in the theory. However, one can remove such modes in a Lorentz-covariant way using the Gupta-Blueler approach \cite{gupta1950theory, bleuler1950neue}. For our proposal, the manifest Lorentz covariance can be dropped out, of choosing the Coulomb gauge. In this situation, the
wave equation of the electromagnetic potential
 can be written as
\begin{equation}
    \left(\frac{1}{c^2}\frac{\partial^2}{\partial t ^2} - \Delta\right)\mathbf{A}(t,\x)=0,
\end{equation}
where $\mathbf{A}$ is the vector potential, and $\Delta$ is the Laplace operator in $\mathbb{R}^d$. From that one can proceed with the usual process of quantization, i.e., define the canonical momentum and impose the commutation relations \cite{kallen2013quantum}.

To proceed we will first restrict the system to a box with sides $(L_1, L_2, \dots, L_{d-2}, a ,b)$ with Dirichlet boundary conditions. Now we have a pure point spectrum and one can go and define the occupation number and a Fock representation. States of such a system are characterized by occupation numbers and, in particular, one can define a vacuum state.

Due to the spectral theorem, the expected value of the Hamiltonian in the vacuum state, that will be denoted by $E_d(L_1, L_2, \dots, L_{d-2}, a ,b)$, is given by the Riemann-Stieljes integral of the spectral measure of the Hamiltonian operator in such a domain. Assuming that $L_i \gg a,b$ for all $i= 1, \dots, d-2$ one can write such a quantity as
%Dirichlet
\begin{equation}\label{eq: Enperf}
E_{d}(L_{1},...,L_{d-2},a,b)=\frac{A_{d-2}}{(2 \pi)^{d-2}} \int\mathrm{d}^{d-2}q \sum_{m,n=1}^{\infty}\hbar\,\omega_{mn}(q)
\end{equation}
%Perióico
%\begin{eqnarray}\label{eq: Enperf}
%&\,&E_{d}(L_{1},...,L_{d-2},a,b)=\frac{A_{d-2}}{(2 \pi)^{d-2}}\nonumber\\
%&\times& \int\mathrm{d}^{d-2}q \left[ \sum_{m,n=1}^{\infty}\omega_{mn}(q) + \sum_{n=1}^{\infty}\frac{\omega_{0n}(q)}{2} + \sum_{m=1}^{\infty}\frac{\omega_{m0}(q)}{2}\right],\nonumber \\
%\end{eqnarray}
where we have used the following definition of the hyperarea
\begin{equation}
    A_{d-2} = \prod_{i=1}^{d-2}L_{i},
\end{equation}
with the frequencies given by
\begin{equation}
    \omega_{mn}(q) = c\sqrt{q^2 +\left(\frac{m\pi }{a} \right)^{2} +\left(\frac{n\pi }{b} \right)^{2}},
\end{equation}
and the continuous momenta
\begin{equation}
    q^2 = q_{1}^{2}+...+q_{d-2}^{2}.
\end{equation}
Of course, as usual in this type of calculation, %So to get some physical observable 
 one need to implement some regularization procedure. %There is many ways to do that, 
Here we will apply an analytic regularization procedure in which we will insert a parameter, $s\in \mathbb{C}$, and our Casimir energy will be given by an analytic extension. With some straightforward manipulations and inserting the parameter in which we will perform the analytic continuation, one can recast the Eq. (\ref{eq: Enperf}), in a more enlightening form
\begin{align}
   &\epsilon_d (a,b; s) = \frac{1}{A_{d-2}}E_{d}(L_{1},...,L_{d-2},a,b; s)
   \nonumber\\
   &
   = \frac{\hbar}{2(2 \pi)^{d-2}}\int\mathrm{d}^{d-2}q \left[ \sideset{}{'}\sum_{m,n=-\infty}^{\infty}\omega_{mn}^{-s}(q)
   \right.\nonumber\\
   &
   \left. -2 \sum_{n=1}^{\infty}\omega_{0n}^{-s}(q) -2 \sum_{m=1}^{\infty}\omega_{m0}^{-s}(q)\right],
\end{align}
where the prime over the summation sign means that the term with $m=n=0$ is removed from the double series. In order to keep with the calculations, we can perform a change of variables in the continuum momenta to a spherical system of coordinates, with radii variable given by $q$ and angular element $\mathrm{d}\Omega_{\mathrm{d}-2}$. The angular integration leads to the factor
\begin{equation}
    \int \mathrm{d} \Omega_{d-2} = \frac{ 2\pi^{\frac{d-2}{2}}}{\Gamma\left(\frac{d-2}{2}\right)},
    \label{solid angle}
\end{equation}
the integration over $q$ can also be performed, leading to
\begin{widetext}
\begin{eqnarray}\label{eq:enzeta}
     \epsilon_d (a,b; s) &=& \frac{\hbar\,c\,\pi^-\frac{d}{2}\Gamma\left(1+\frac{s}{2}-\frac{d}{2}\right)}{2^{d-1}\Gamma\left(\frac{s}{2}\right)}\left[Z_2\left(\frac{1}{a}, \frac{1}{b}; s-d+2\right) - \left(\frac{1}{a^{d-2-s}} + \frac{1}{b^{d-2-s}}\right)\zeta(s-d+2)\right], 
\end{eqnarray}
\end{widetext}
where we have used the definition of the Epstein zeta-function
\begin{equation}\label{eq:eps}
    Z_2(x,y;s) = \sideset{}{'}\sum_{m,n=-\infty}^{\infty} \left[(xm)^2 + (yn)^2 \right]^{-s/2},
\end{equation}
and the definition of the Riemann zeta-function
\begin{equation}
    \zeta(s) = \sum_{n=1}^{\infty}\frac{1}{n^s}.
\end{equation}
%The zero-point energy given at Eq. (\ref{eq:enzeta}) \textcolor{blue}{is} still divergent for $s=-1$ and $d\geq 3$. 
%However the analytic extension of the Epstein zeta-function is well-known, as well the analytic extension of the Riemann zeta-function. In particular we will use the equations known as reflection formula, for both of them. In our case, one can 
Using the reflection formula, we can write the Epstein zeta-function as
\begin{eqnarray}
  Z_2\left(\frac{1}{a}, \frac{1}{b}; s-d+2\right) &=&\frac{ab}{\pi{^{s+\frac{d-1}{2}}}} \frac{\Gamma\left(\frac{d-s}{2}\right)}{\Gamma\left(\frac{2-d+s}{2}\right)}\nonumber \\
  &\times& Z_2(a,b;d-s),
\end{eqnarray}
to the Riemann zeta-function, we use the reflection formula alongside with the Legendre duplication formula for the $\Gamma$-function to obtain
\begin{equation}\label{eq:rie}
    \Gamma\left(1+\frac{s}{2}-\frac{d}{2}\right)\zeta(s-d+2) =\frac{\Gamma\left(\frac{d-1-s}{2}\right)}{\pi^{-s+\frac{d}{2}+1}}\zeta(d-1-s).
\end{equation}
One can write general expression for $ \epsilon_d (a,b; s)$ employing analytic extension procedure. The Casimir energy for the waveguide is obtained for $s=-1$.
%\begin{eqnarray}\label{eq:enzeta2}
 %    \epsilon_d (a,b; -1) &=& -\frac{\pi^{-\frac{d +1}{2}}}{2^{d}}\left[ab\pi^{1-\frac{d-1}{2}}\Gamma\left(\frac{d+1}{2}\right) Z_2\left(a, b; d+1\right)- \left(\frac{1}{a^{d-1}} + \frac{1}{b^{d-1}}\right)\frac{\Gamma\left(\frac{d}{2}\right)}{\pi^{\frac{d}{2} +2}}\zeta(d)\right] \nonumber \\
%\end{eqnarray}
Which is finite for $d\geq 3$. In our case of interest, $d=3$, one can write 
\begin{equation}
    \epsilon_3 (a,b; -1) =  \frac{\hbar\,c}{16\pi^2}\left(\frac{1}{a^{2}} + \frac{1}{b^{2}}\right)\zeta(3)-\frac{\hbar\,c}{8\pi}ab\,Z_2\left(a, b; 4\right).
    \label{perfect ZP energy}
\end{equation}
The sign difference between the two contributions is what ensures the well known behavior of change of sign of the Casimir force in a box geometry.

%Analytic continuation can be also be used to obtain the Casimir force of imperfect conductors. Recently Ref. \cite{Arias:2023vfr} propose the use of approximate functional equations in order to obtain corrections of finite conductivity for the Casimir force. 

In Section \ref{sec:stattionary modes} we explore the stationary modes in a dielectric rectangular waveguide. Later, such modes are being used alongside with the analytic continuation procedure and plasma frequencies definition to obtain the Casimir force of such a system. 
%In the Section \ref{sec:casimir energy} we employ a different method of regularization to obtain corrections to the Casimir force of a rectangular waveguide with finite conductivity.

\section{Stationary electromagnetic modes in dielectric waveguide}
\label{sec:stattionary modes}
\begin{figure}[h!]
\begin{center}
\includegraphics[scale=0.25]{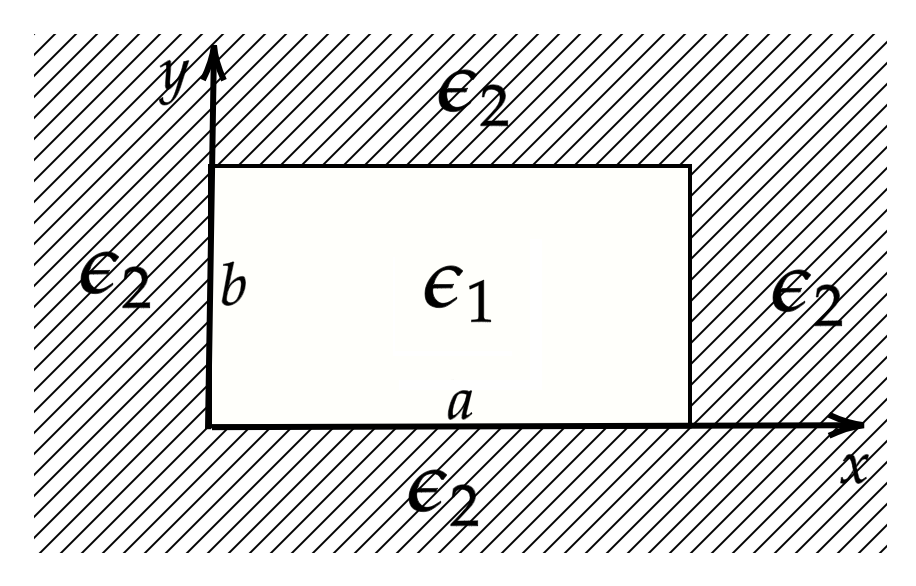}\\
\caption{Geometry of the electromagnetic waveguide with rectangular cross-section and filled with different dielectric materials from inside and outside.} \label{box}
\end{center}
\end{figure}

In order to find the zero-point energy of the electromagnetic field in a waveguide fulfill with dielectric material, we find the surface stationary modes in this specific geometry.
In this manner, let us begin the discussion by considering the Maxwell equations for dielectric (non-magnetic) materials
\begin{align}
  \nabla\cdot{\bf D}&=0,\nonumber\\
  \nabla\cdot{\bf B}&=0,\nonumber\\
  \nabla\times{\bf E}&=-\frac{1}{c}\frac{\partial{\bf B}}{\partial t},\nonumber\\
    \nabla\times{\bf B}&=\frac{1}{c}\frac{\partial{\bf D}}{\partial t},
\end{align}
where the electric displacement vector is given by ${\bf D}({\bf r},t)=\epsilon(\omega){\bf E}({\bf r},t)$. Since we are assuming an isotropic homogeneous media, the dielectric constant is not dependent on the spatial coordinates, but it depends on the wave frequency. We are interested in finding stationary solutions of the field equations of the form
\begin{align}
   {\bf E}({\bf r},t)&={\bf E}_0({\bf r})e^{-i\omega t},\nonumber\\
   {\bf B}({\bf r},t)&={\bf B}_0({\bf r})e^{-i\omega t}.
\end{align}
Our purpose is to find the natural frequencies $\omega$ that satisfy the Maxwell equations with the appropriate boundary conditions. Let us consider a geometry given by a waveguide along the $z$-direction that have $a$ length in the $x$-direction and $b$ length in the $y$-direction, that is, the waveguide is defined by the set of ordered pairs $\{(x,y);\, x\in[0,a],\, y\in[0,b]\}$. The non-perfect conducting materials have dielectric constants $\epsilon_1$ inside and $\epsilon_2$ outside the waveguide, see Figure (\ref{box}).

By the translation symmetry along the $z$-axis we expect that the electric and magnetic fields only depend on coordinates $x$ and $y$, while in the $z$ direction we expect a free plane wave term. Hence, we look for a solution of the form
\begin{align}
    {\bf E}_0({\bf r})&=\left(e_x(x,y)\hat{i}+e_y(x,y)\hat{j}+e_z(x,y)\hat{k}\right)e^{ikz},\nonumber\\
    {\bf B}_0({\bf r})&=\left(b_x(x,y)\hat{i}+b_y(x,y)\hat{j}+b_z(x,y)\hat{k}\right)e^{ikz}.
    \label{E0B0}
\end{align}
By using the Gauss law in the electric displacement field, one find that
\begin{equation}
    \frac{\partial e_x}{\partial x}+\frac{\partial e_y}{\partial y}+ik\,e_z=0.
    \label{divE}
\end{equation}
From the Faraday law we have that the spatial dependence of the magnetic field is ${\bf B}_0=-i(c/\omega)\nabla\times{\bf E}_0$, and therefore we can write
\begin{align}
    {\bf B}_0=-i\left(\frac{c}{\omega}\right)\bigg\{&\left(\frac{\partial e_z}{\partial y}-ike_y\right)\hat{i}+\left(ike_x-\frac{\partial e_z}{\partial x}\right)\hat{j}\nonumber\\
    &+\left(\frac{\partial e_y}{\partial x}-\frac{\partial e_x}{\partial y}\right)\hat{k}\bigg\}e^{ikz},
\end{align}
from this equation it follows immediately that the Gauss law for the magnetic field is satisfied, i.e. $\nabla\cdot{\bf B}_0=0$. Also, after some manipulations,  
%from the above relation 
we have straightforward that $\nabla\times{\bf B}_0=-i(c/\omega)\nabla\times(\nabla\times{\bf E}_0)=i(c/\omega)\nabla^2{\bf E}_0$. Replacing this into the Ampere law, we obtain the wave equation
\begin{equation}
    \nabla^2{\bf E}_0+\frac{\omega^2}{c^2}\epsilon(\omega){\bf E}_0=0.
    \label{wave eq}
\end{equation}
By using Eq. (\ref{E0B0}) into Eq. (\ref{wave eq}), we have that the components of the electric field must satisfy 
\begin{equation}
    \frac{\partial^2 e_i}{\partial x^2}+\frac{\partial^2 e_i}{\partial y^2}-K^2\,e_i=0,
    \label{we}
\end{equation}
where the index is $i=\{x,y,z\}$ and we have defined the wave number on the transverse section of the waveguide
\begin{equation}
   K^2=k^2-\frac{\omega^2}{c^2}\epsilon(\omega).
\end{equation}
%The appropriate boundary conditions for the rectangular waveguide are discussed in the Appendix \ref{app:A}.
%There we conclude that the components of the electric or magnetic fields must satisfy specific continuity conditions at the interfaces between dielectric.

%\subsection{Stationary modes}
From the analysis of the boundary conditions, discussed in Appendix \ref{app:A}, we have that there are some conditions that are incompatible.
The origin of this problematic is the impossibility of defining a normal and tangential component in the corners of the rectangular waveguide. For example, the $y$-component of the electric field is normal to the horizontal surfaces 
horizontal surfaces $H_1 = \{(x,y) \in [0,a]\times[0,0]\}$ and $H_2 = \{(x,y)\in[0,a]\times[b,b]\}$,
%$y=\{0,b\}$
so that it must be satisfied that $\epsilon(\omega)\,e_y$ should be continuous at that surfaces, however the same $y$-direction is tangential when referring to the vertical surfaces %$x=\{0,a\}$ 
$V_1 = \{(x,y) \in [0,0]\times[0,b]\}$   and $V_2 =  \{(x,y) \in [a,a]\times[0,b]\}$,
where only $e_y$ must be continuous. For this case one can conclude that, in order to satisfy both conditions at the four corners of the waveguide, $(x,y)\in\{(0,0),(a,0),(0,b),(a,b)\}$, it must be that the transverse component should vanish, i.e., $e_x=e_y=0$.
We can extend this observation and look for stationary solutions where we have always $e_y(x,y)\equiv 0$, for all the points of space and this conditions will define our $X$-mode solution. Similarly, we can search for and independently solutions where $e_x(x,y)\equiv 0$, and this conditions will define our $Y$-mode.

%\subsubsection{Surface stationary X-modes}

First, let us discuss the surface stationary $X$-modes. For these modes, we assume that the electric field only has $e_x$ and $e_z$ components, while we set $e_y=0$. By using the Gauss equation, Eq. (\ref{divE}), we note that the $z$-component is completely determined by the $e_x$ component
\begin{equation}
    e_z=\frac{i}{k}\frac{\partial e_x}{\partial x},
\end{equation}
so that the only degree of freedom is $e_x$. By considering the boundary conditions on the waveguide surfaces we have that for the vertical surfaces $V_1$ and $V_2$: 
%\noindent
%\textit{Vertical surfaces $x=\{0,a\}$}:
the components $\epsilon(\omega)e_x$, $\partial_xe_x$ and $\partial_ye_x$ must be continuous.
For the horizontal surfaces, $H_1$ and $H_2$:
%\noindent
%\textit{Horizontal surfaces $y=\{0,b\}$}:
the components $e_x$, $\partial_xe_x$ and $\partial_ye_x$ must be continuous.

Here we note some apparent contradiction. From the continuity of the normal component of the electric field we have that $\epsilon(\omega)e_x$ must be continuous at the verticals $V_1$ and $V_2$; while from the continuity of the tangential component of the electric field only $e_x$ must be continuous at horizontals $H_1$ and $H_2$. But we know that it is impossible to both $e_x$ and $\epsilon(\omega) e_x$ be continuous on the interface between dielectrics with different properties. In order to fulfill both requirement independently, we have two possibilities, that defines the mode $X1$ solutions and the mode $X2$ solutions.

%\noindent
%\underline{\textit{Mode X1 solution}}:\\
For the mode $X1$ solutions, the components $\epsilon(\omega)e_x$, $\partial_xe_x$ and $\partial_ye_x$ must be continuous. Also, it is required that, at the horizontal surfaces of the waveguide, the transverse component of the electric field is null, i.e, $e_x(x,0)=e_x(x,b)=0$, for all values $0\leq x\leq a$. The wave equation given by Eq. (\ref{we}), can be solve by separation of variables method. 
%\begin{equation}
 %   \frac{\partial^2 e_x}{\partial x^2}+\frac{\partial^2 e_x}{\partial y^2}-K^2\,e_x=0,
  %  \label{we1}
%\end{equation}
%where $K^2=k^2-\frac{\omega^2}{c^2}\epsilon(\omega)$. 
Hence, we look for solutions of the form $e_x(x,y)=f(x)g(y)$ such that these functions satisfy
\begin{equation}
    \frac{f''(x)}{f(x)}+\frac{g''(y)}{g(x)}-K^2=0.
\end{equation}
%\noindent
%\underline{\textit{Dispersion relation  for A1 modes}}:\\
%\noindent
By considering explicitly the boundary conditions for $X1$-modes and avoiding nonphysical exponentially growing solutions, we find oscillatory solutions in the $y$-direction and exponentially decaying functions in the $x$-direction. This means that the $X1$-modes solutions are given by $e^{(n)}_x(x,y)=f_n(x)g_n(y)$ with
\begin{equation}
    g_n(y)=\sin\left(\frac{n\pi}{b}y\right),
\end{equation}
where $n=1,2,3,...$. We have a surface decaying behavior from the waveguide surfaces at the $x$-direction
\begin{equation}
    %\[
    f_n(x)= 
\begin{cases}
    A e^{\Lambda_2 x},& \,x<0\\
    Be^{-\Lambda_1 x}+Ce^{\Lambda_1 x},  &0<x<a\\
    De^{-\Lambda_2 x},& a<x\\
\end{cases}
%\]
\end{equation}
where we have defined the wave numbers
\begin{equation}
    \Lambda_{1,2}=\sqrt{\left(\frac{n\pi}{b}\right)^2+k^2-\frac{\omega^2}{c^2}\epsilon_{1,2}(\omega)}.
    \label{Lambda}
\end{equation}
These wave numbers indeed depend on integer $n$, wave number $k$, and on the wave frequency $\omega$,
this means $\Lambda_{1,2}=\Lambda_{1,2}(n,k,\omega)$. However, we have omitted an explicit dependence on these parameters in order to emphasize the dependence of $\Lambda_{1,2}$ on the different dielectric constants of the materials $\epsilon_{1,2}$. 
%These modes indeed satisfy the condition $e_x(x,0)=e_x(x,b)=0$, for all values $0<x<a$. 
Now, considering the continuity of $\epsilon(\omega)e_x$ and the derivative $\partial_xe_x$ at verticals $V_1$ and $V_2$, we obtain the following condition in order to obtain non-trivial solutions for the $X1$-modes
\begin{equation}
    {\cal F}_{X1}(n, k, \omega)=e^{a\Lambda_1}\left(\frac{\Lambda_1\epsilon_2+\Lambda_2\epsilon_1}{\Lambda_1\epsilon_2-\Lambda_2\epsilon_1}\right)^2-e^{-a\Lambda_1}=0.
    \label{FA1}
\end{equation}
This equation give us all the values of the allowed frequencies $\omega$ that contribute to  the zero-point energy.

%\noindent
%\\
%\underline{\textit{Mode X2 solution}:}\\

Let us now discuss the $X2$-mode solutions. Now, the components $e_x$, $\partial_xe_x$ and $\partial_ye_x$ must be continuous, and the transverse electric field must vanishes at the vertical surfaces of the waveguide, i.e.,  $e_x(0,y)=e_x(a,y)=0$, for all values $0\leq y\leq b$.
As before, we look for solutions that are exponentially decaying from the waveguide surface.
By using the separation of variables method, we note that it must be referred to oscillatory solutions in the $x$-direction and exponentially decaying solutions in the $y$-direction. For the $X2$-modes we have solutions of the form $e_x^{(n)}(x,y)=f_n(x)g_n(x)$, where in this case
\begin{equation}
    f_n(x)=\sin\left(\frac{n\pi}{a}x\right),
\end{equation}
where $n=1,2,3,...$ and in the $y$-direction it is found an exponential decaying behavior
\begin{equation}
    %\[
    g_n(y)= 
\begin{cases}
    {\cal A} e^{\Upsilon_2 y},&\, y<0\\
    {\cal B}e^{-\Upsilon_1 y}+\tilde{C}e^{\Upsilon_1 y}, & 0<y<b\\
    {\cal D}e^{-\Upsilon_2 y},& b<y\\
\end{cases}
%\]
\end{equation}
where we have defined the wave numbers
\begin{equation}
    \Upsilon_{1,2}=\sqrt{\left(\frac{n\pi}{a}\right)^2+k^2-\frac{\omega^2}{c^2}\epsilon_{1,2}(\omega)},
    \label{Upsilon}
\end{equation}
here we have that $\Upsilon_{1,2}=\Upsilon_{1,2}(n,k,\omega)$. Now, considering that for the $X2$-modes we need to ensure the continuity of $e_x(x,y)$ and of $\partial_ye_x(x,y)$ we find the condition to have non-trivial solutions for the $X2$-mode as 
\begin{equation}
    {\cal F}_{X2}(n,k,\omega)=e^{b\Upsilon_1}\left(\frac{\Upsilon_1+\Upsilon_2}{\Upsilon_1-\Upsilon_2}\right)^2-e^{-b\Upsilon_1}=0,
    \label{FA2}
\end{equation}
this conditions give us some normal frequencies that contribute to the zero-point energy.

%\subsubsection{Surface stationary $Y$-modes}
Now let us focus in the surface stationary $Y$-modes, in this case we look for stationary field solutions where the electric field that has only $e_y$ and $e_z$ components, while $e_x=0$. By using the Gauss law Eq. (\ref{divE}) we see that the $e_z$ is not independent and therefore, $e_y$ is the only degree of freedom for the $Y$-modes. In turn, this component must satisfy the boundary conditions at the waveguide interface between the dielectrics. For the vertical surfaces, $V_1$ and $V_2$:
%
%\noindent
%\textit{Vertical surfaces $x=\{0,a\}$}: 
the components $e_y$, $\partial_xe_y$ and $\partial_ye_y$ must be continuous.
While for the horizontal surfaces $H_1$ and $H_2$:
%\noindent
%\textit{Horizontal surfaces $y=\{0,b\}$}: 
the components $\epsilon(\omega)e_y$, $\partial_xe_y$ and $\partial_ye_y$ must be continuous.
In order to accomplish this, we need to add some supplementary conditions that define two kinds of modes for the $Y$-solutions.

%\noindent
%\\
%\underline{\textit{Mode Y1 solution}}:\\
Analogously to the $X$-modes case, the mode $Y1$ solution will be given by
the components $e_y$, $\partial_xe_y$ and $\partial_ye_y$ must be continuous with the supplementary condition that 
the transverse component of the electric field 
 is null at the horizontal surfaces of the waveguide, i.e., $e_y(x,0)=e_y(x,b)=0$, for all values $0\leq x\leq a$.
 By using this conditions and looking for non-null solutions one obtains an equation for the frequencies of the electromagnetic wave inside the waveguide. 
\begin{equation}
    {\cal F}_{Y1}(n,k,\omega)=e^{a\Lambda_1}\left(\frac{\Lambda_1+\Lambda_2}{\Lambda_1-\Lambda_2}\right)^2-e^{-a\Lambda_1}=0.
    \label{FB1}
\end{equation}
%\noindent
%\underline{\textit{Mode Y2 solution}}:\\

Finally, for the mode $Y2$ solution,
the components $\epsilon(\omega)e_y$, $\partial_xe_y$ and $\partial_ye_y$ must be continuous with the additional condition $e_y(0,y)=e_y(a,y)=0$, for all values $0\leq y\leq b$.
In this case, we find that the condition that defines the frequency for the surface modes of kind $Y2$ is given by
\begin{equation}
    {\cal F}_{Y2}(n,k,\omega)=e^{b\Upsilon_1}\left(\frac{\Upsilon_1\epsilon_2+\Upsilon_2\epsilon_1}{\Upsilon_1\epsilon_2-\Upsilon_2\epsilon_1}\right)^2-e^{-b\Upsilon_1}=0,
    \label{FB2}
\end{equation}
where $\Lambda_{1,2}$ and $\Upsilon_{1,2}$ have being defined in Eq. (\ref{Lambda}) and Eq. (\ref{Upsilon}). The principal result of this Section is given by the conditions given by Eq. (\ref{FA1}) and Eqs.~(\ref{FA2}-\ref{FB2}). These expressions give us all the possible values of the surface mode frequencies for the electromagnetic field inside the dielectric cavity. With this result, we are able to calculate the zero-point energy inside the waveguide and consequently analyze the Casimir effect.

\section{Casimir energy in dielectric waveguide}
\label{sec:casimir energy}
In the previous section, we have focus on the exponentially decaying stationary solutions of the electromagnetic field in the waveguide geometry. These are referred as surface modes, and  it can be proved that, in the small distances, only these modes contribute to vacuum energy \cite{langbein1973macroscopic, barton1979some}. The zero-point vacuum energy is given by the sum of all the possible frequency that ensure the existence of modes $X$ and $Y$. Then in the case of dielectric waveguide consider here we zero-point energy is
\begin{equation}
    E_{ZP}=\sum_N\frac{1}{2}\hbar\omega_N^X+\sum_N\frac{1}{2}\hbar\omega_N^Y,
\end{equation}
where $\omega_N^X$ and $\omega_N^Y$ are the allowed frequencies for $X$-mode and $Y$-mode, respectively. Since these modes depend on the continuum variable $k$, associated with the plane waves in the $z$-direction and on the discrete index $n$, that ensure the appropriate boundary conditions the sum refers to
$$\sum_N\rightarrow\left(\frac{L_z}{2\pi}\right)\int_{-\infty}^{\infty}dk\sum_{n=1}^\infty\sum_r,$$
where $L_z$ is the dimension of the large non-compactified dimension of the waveguide and the sum $\sum_r$ refers to roots or zeros of the boundary condition equations given by Eq.(\ref{FA1}) and Eqs. (\ref{FA2}-\ref{FB2}). For example, in order to consider an $X$-mode frequency it must be ensured that this frequency is a root of the functions that defines the boundary conditions, i.e., ${\cal F}_{X}(n,k,\omega_r^X)=0$, where we must consider the function indicated in Eq. (\ref{FA1}) or  Eq. (\ref{FA2}). It is clear that the frequencies that satisfy this conditions depends on $n$ and $k$, such that $\omega_r^X=\omega_r^X(n,k)$ and the zero-point energy can be written as
\begin{align}
    \!\!\!\!E_{ZP}=\left(\frac{\hbar L_z}{4\pi}\right)\int_{-\infty}^{\infty}\!\!\!\!\!\!\!dk\sum_{n=1}^\infty\sum_r\bigg(\omega_r^X(n,k)+\omega_r^Y(n,k)\bigg).
\end{align}
To proceed, we use the argument theorem of complex variables. This theorem indicates that for any meromorphic complex function, it is satisfied that
\begin{equation}
    \frac{1}{2\pi i}\oint_{C}\frac{f'(z)}{f(z)}dz={\cal N}-{\cal P},
\end{equation}
where $C$ is any closed contour, ${\cal N}$ is the quantity of roots or zeros of the function inside $C$ and ${\cal P}$ is the number of poles inside $C$. This theorem could be extended in such a way that
\begin{equation}
    \frac{1}{2\pi i}\oint_{C}z\frac{f'(z)}{f(z)}dz=\sum_rz_r-\sum_pz_p,
\end{equation}
where the first sum is restricted to the roots of the function inside the contour, $f(z_r)=0$, while the second sum is for the poles of the function, $f(z_p)\rightarrow\infty$.
%this means that the integral above give directly the difference between the sum of all the function roots $z_r$,  and the sum of all the function poles, $z_p$. 
Since $\omega_r^X$ are the roots of the function ${\cal F}_X(n,k,\omega)$, for a given $n$ and $k$, one can write the sum as
\begin{equation}
    \sum_r\omega_r^X
    =\frac{1}{2\pi i}\oint_{C}\omega\frac{{\cal F'}_X(\omega)}{{\cal F}_X(\omega)}d\omega+
    \sum_p\omega_p^X,
    \nonumber
\end{equation}
where $\omega_p^X$ are poles of ${\cal F}_X(\omega)$
and we consider the contour $C$ as a semicircle (to be considered of large radius) that contains the imaginary axis and is closed counterclockwise from the right. The last sum  in the above equation is the sum of all the poles of the function, and we assume that this does not depend on the boundaries in such a way that would not contribute to the Casimir energy. Hence, we obtain that
\begin{widetext}
\begin{align}
    E_{ZP}=\left(\frac{\hbar L_z}{4\pi}\right)\frac{1}{2\pi i}\int_{-\infty}^{\infty}dk\sum_{n=1}^\infty\bigg(\oint_{C}\omega\frac{{\cal F'}_X(n,k,\omega)}{{\cal F}_X(n,k,\omega)}d\omega
    +\oint_{C}\omega\frac{{\cal F'}_Y(n,k,\omega)}{{\cal F}_Y(n,k,\omega)}d\omega\bigg),    
\end{align}
%\end{widetext}
here the notation is such that the derivative ${\cal F'}_X(n,k,\omega)=\partial{\cal F}_X(n,k,\omega)/\partial\omega$. In the limit of infinite radius for the contour $C$ the only non-zero contribution to the complex integral above comes from the imaginary axis $\omega=i\xi$ where $\xi\in(+\infty,-\infty)$. By a change of variables, we define $F_X(\xi)={\cal F}_X(i\xi)$.
Performing an integration by parts and some manipulations, we find
%\begin{equation}
%    E(a,b)=\left(\frac{\hbar L_z}{8\pi^2}\right)\int_{-\infty}^{\infty}dk\sum_{n=1}^\infty\left(\int_{-\infty}^\infty d\xi\log F_A^{(n)}(\xi) +
%    \int_{-\infty}^\infty d\xi\log F_B^{(n)}(\xi) \right),
%\end{equation}
\begin{align}
    E_{ZP}=\left(\frac{\hbar L_z}{8\pi^2}\right)\int_{-\infty}^{\infty}dk\sum_{n=1}^\infty
    \int_{-\infty}^\infty d\xi
    \bigg(\ln F_{X}(n,k,\xi) +
    %\ln F_{X2}(n,k,\xi) 
    %+\ln F_{Y1}(n,k,\xi) 
    \ln F_{Y}(n,k,\xi)
    \bigg).
\end{align}
%\begin{align}
 %   E_{ZP}=\left(\frac{\hbar L_z}{8\pi^2}\right)\int_{-\infty}^{\infty}dk\sum_{n=1}^\infty
  %  \bigg(&\int_{-\infty}^\infty d\xi\ln F_{A1}^{(n)}(\xi) +
   % \int_{-\infty}^\infty d\xi\ln F_{A2}^{(n)}(\xi) 
    %+\int_{-\infty}^\infty d\xi\ln F_{B1}^{(n)}(\xi) +
    %\int_{-\infty}^\infty d\xi\ln F_{B2}^{(n)}(\xi)
    %\bigg),
%\end{align}
Now, we have to consider that there are two contributions for each $X$-mode and $Y$-mode. Hence, by using explicitly the boundary equations for all the surface modes Eq. (\ref{FA1}) and Eqs. (\ref{FA2}-\ref{FB2}) we obtain the zero-point energy in the dielectric waveguide as
%\begin{widetext}

\begin{align}
    E_{ZP}=\left(\frac{\hbar L_z}{8\pi^2}\right)\int_{-\infty}^{\infty}dk\sum_{n=1}^\infty\int_{-\infty}^\infty d\xi&
    \bigg\{\ln\left(e^{a{\lambda}_1}\left(\frac{{\lambda}_1\epsilon_2+{\lambda}_2\epsilon_1}{{\lambda}_1\epsilon_2-{\lambda}_2\epsilon_1}\right)^2-e^{-a{\lambda}_1}\right)
    +
   \ln\left(e^{b{\upsilon}_1}\left(\frac{{\upsilon}_1+{\upsilon}_2}{{\upsilon}_1-{\upsilon}_2}\right)^2-e^{-b{\upsilon}_1}\right)\nonumber\\ 
    &+\ln\left(e^{b{\upsilon}_1}\left(\frac{{\upsilon}_1\epsilon_2+{\upsilon}_2\epsilon_1}{{\upsilon}_1\epsilon_2-{\upsilon}_2\epsilon_1}\right)^2-e^{-b{\upsilon}_1}\right)
    +
   \ln\left(e^{a{\lambda}_1}\left(\frac{{\lambda}_1+{\lambda}_2}{{\lambda}_1-{\lambda}_2}\right)^2-e^{-a{\lambda}_1}\right)
    \bigg\},
    \label{ZP}    
\end{align}
\end{widetext}
%\newpage
%\begin{align}
%    E(a,b)=\left(\frac{\hbar L_z}{8\pi^2}\right)\int_{-\infty}^{\infty}dk\sum_{n=1}^\infty
%    \bigg\{&\int_{-\infty}^\infty d\xi\log\left(\left(\frac{\tilde{\gamma}_1\epsilon_2+\tilde{\gamma}_2\epsilon_1}{\tilde{\gamma}_1\epsilon_2-\tilde{\gamma}_2\epsilon_1}\right)^2e^{2a\tilde{\gamma}_1}-1\right)\nonumber\\ 
%    +
%    &\int_{-\infty}^\infty d\xi\log\left(\left(\frac{\tilde{\lambda}_1+\tilde{\lambda}_2}{\tilde{\lambda}_1-\tilde{\lambda}_2}\right)^2e^{2b\tilde{\lambda}_1}-1\right) \nonumber\\
%    +&\int_{-\infty}^\infty d\xi\log\left(\left(\frac{\tilde{\lambda}_1\epsilon_2+\tilde{\lambda}_2\epsilon_1}{\tilde{\lambda}_1\epsilon_2-\tilde{\lambda}_2\epsilon_1}\right)^2e^{2b\tilde{\lambda}_1}-1\right)\nonumber\\ 
 %   +
 %   &\int_{-\infty}^\infty d\xi\log\left(\left(\frac{\tilde{\gamma}_1+\tilde{\gamma}_2}{\tilde{\gamma}_1-\tilde{\gamma}_2}\right)^2e^{2a\tilde{\gamma}_1}-1\right)
 %   \bigg\},
  %  \label{Ecasimir}
%\end{align}
where we have denoted the wave numbers for imaginary frequencies by
\begin{align}\label{lambda}
{\lambda}_{1,2}&=\sqrt{\left(\frac{n\pi}{b}\right)^2+k^2+\frac{\xi^2}{c^2}\epsilon_{1,2}(i\xi)},
\end{align}
and
\begin{align}
{\upsilon}_{1,2}&=\sqrt{\left(\frac{n\pi}{a}\right)^2+k^2+\frac{\xi^2}{c^2}\epsilon_{1,2}(i\xi)},
    \label{upsilon}
\end{align}
these variables depend on the integer $n$, the wave number in the $z$-direction, $k$, and on the imaginary frequency $\xi$. It is worth to note that in general the dielectric constant depend on the wave frequency. 
%This depends introduce the plasma frequency that characterize the material. 

%\subsection{Parallel dielectric plates limit}
The main result of this work is Eq. (\ref{ZP}), that gives the zero-point energy associated to an electromagnetic field in a waveguide of rectangular cross-section surrounded by two dielectric materials. In our case the finite dimensions of the waveguide are $a, b$ while the third dimension is $L_z$ and $L_z\gg a, b$.
The original Lifshitz result, in the case of two parallel plates, can be recovered from our result 
 if we keep one direction of the waveguide fixed and allow the other ones goes to infinity.
%if we consider one of the finite dimensions of the waveguide is finite while the other tends to infinity. 
More specific, in this limit we can consider that $a$ is finite while  $L_z, b\gg a$.
In order to take this limit, we use the following relation
$$\lim_{b\rightarrow\infty}\sum_{n=1}^{\infty}f\left(\frac{n\pi}{b}\right)=\left(\frac{b}{2\pi}\right)\int_{-\infty}^{\infty}d\tilde{k}\,f(\tilde{k})$$
and define the variable
$$  K_i=\sqrt{\tilde{k}^2+k^2+\frac{\xi^2}{c^2}\epsilon_i(i\xi)}.$$
In this manner, we can write 
\begin{align}
    E^{plates}_{ZP}=&\,{\cal N}
    \!\!\int\!dk d\tilde{k}\,d\xi
    %\int_{-\infty}^{\infty}dk&
    %\left(\frac{b}{2\pi}\right)
    %\int_{-\infty}^{\infty}d\tilde{k}
    %\int_{-\infty}^\infty d\xi
    \bigg\{
    \ln\left(\left(\frac{{K}_1+{K}_2}{{K}_1-{K}_2}\right)^2e^{2a{K}_1}-1\right)
    \nonumber\\
        &
    +
    \ln\left(\left(\frac{{K}_1\epsilon_2+{K}_2\epsilon_1}{{K}_1\epsilon_2-{K}_2\epsilon_1}\right)^2e^{2a{K}_1}-1\right)\bigg\}.       
    \label{ZPplates}
\end{align}
where
${\cal N}=\hbar L_z b/16\pi^3.$
This expression is obtained by considering the limits of the contributions of the first and fourth terms inside the integral of the general result Eq. (\ref{ZP}). This is because the dependence of second and third term inside the integral Eq. (\ref{ZP}) only depends on length $b$ through the exponential factor that give us an infinite (constant) term in the limit $b\rightarrow\infty$.
In the above integral, we have that all the limits of integration are from $-\infty$ to $+\infty$. By considering $k$ and $\tilde{k}$ as coordinates of a two-dimensional space, defining $\kappa=\sqrt{k^2+\tilde{k}^2}$ and performing the angular integration we obtain
\begin{align}
   \!\! E^{plates}_{ZP}=&{\tilde{\cal N}}
    %\int dk\,d\tilde{k}d\xi
    \int_{0}^{\infty}\!\! \!\!\!d\kappa\,\kappa
    %\left(\frac{b}{2\pi}\right)
    %\int_{-\infty}^{\infty}d\tilde{k}
    \int_{-\infty}^\infty \!\!\!\!\!d\xi
    %\nonumber\\    &
    \bigg\{%\nonumber\\
        \ln\left(\left(\frac{{K}_1+{K}_2}{{K}_1-{K}_2}\right)^2e^{2a{K}_1}-1\right)
    \nonumber\\&
    +
    \ln\left(\left(\frac{{K}_1\epsilon_2+{K}_2\epsilon_1}{{K}_1\epsilon_2-{K}_2\epsilon_1}\right)^2e^{2a{K}_1}-1\right)\bigg\},     
    \label{ZPplatesMILONI}
\end{align}
where
$ {\tilde {\cal N}} = \hbar L_z b/8\pi^2$.
The above equation is exactly the Lifshitz formula, for the case of two parallel plates (separated by a finite distance $a$) with a media of dielectric constant  $\epsilon_1$ between two media of dielectric constant $\epsilon_2$ in a slab configuration.

%\subsection{Perfectly conducting waveguide limit}

Let us discuss the general result given by Eq. (\ref{ZP})
%i.e, about the Casimir energy of the dielectric waveguide 
for the case where the surfaces are perfect conductors.
%In this section we obtain the expression of the zero-point energy for the case of ideal perfecting surfaces in the waveguide as a limiting case of the general expression obtained in this work, see Eq. (\ref{ZP}).
In order to do this, first one can rewrite the expression for the zero-point energy in a more compact way
\begin{align}
    E_{ZP}=\left(\frac{\hbar L_z}{8\pi^2}\right)&\int_{-\infty}^{\infty}dk\sum_{n=1}^\infty\int_{-\infty}^\infty d\xi
    \bigg\{\ln\left({\cal X}_1e^{a{\lambda}_1}-e^{-a{\lambda}_1}\right)
    \nonumber\\
    &\!\!\!\!\!\!\!+   \ln\left({\cal X}_2e^{b{\upsilon}_1}-e^{-b{\upsilon}_1}\right)
    +\ln\left({\cal Y}_1e^{b{\upsilon}_1}-e^{-b{\upsilon}_1}\right)\nonumber\\
    &\!\!\!\!\!\!\!+
   \ln\left({\cal Y}_2e^{a{\lambda}_1}-e^{-a{\lambda}_1}\right)
    \bigg\},
\end{align}
where the reflectivity indexes are defined as follows
\begin{align}
    {\cal X}_1&=\left(\frac{{\lambda}_1\epsilon_2+{\lambda}_2\epsilon_1}{{\lambda}_1\epsilon_2-{\lambda}_2\epsilon_1}\right)^2,\hspace{.5cm}
    {\cal X}_2=\left(\frac{{\upsilon}_1+{\upsilon}_2}{{\upsilon}_1-{\upsilon}_2}\right)^2,\nonumber\\
    {\cal Y}_1&=\left(\frac{{\upsilon}_1\epsilon_2+{\upsilon}_2\epsilon_1}{{\upsilon}_1\epsilon_2-{\upsilon}_2\epsilon_1}\right)^2,\hspace{.5cm}
    {\cal Y}_2=\left(\frac{{\lambda}_1+{\lambda}_2}{{\lambda}_1-{\lambda}_2}\right)^2.
    \label{reflex}
\end{align}

In the ideal case the media inside the waveguide is a perfect vacuum with $\epsilon_1=1$ whereas the boundaries are perfect reflecting surfaces with $\epsilon_2\rightarrow\infty$.
%In the ideal case of perfect conducting surface we also consider vacuum inside the waveguide with dielectric constant $\epsilon_1=1$. 
From this follows  that the wave numbers inside the waveguide are given by $\lambda_1\rightarrow\lambda^{(0)}$ and $\upsilon_1\rightarrow\upsilon^{(0)}$ with
\begin{align}
    {\lambda}^{(0)}&=\sqrt{\left(\frac{n\pi}{b}\right)^2+k^2+\frac{\xi^2}{c^2}},\nonumber\\
    {\upsilon}^{(0)}&=\sqrt{\left(\frac{n\pi}{a}\right)^2+k^2+\frac{\xi^2}{c^2}}.
    \label{lambda upsilon 0}
\end{align}
For the other side, outside the waveguide one has a perfect conductor with $\epsilon_2\rightarrow\infty$ and  consequently $\lambda_2, \upsilon_2\rightarrow\infty$. 
It can be shown that in this limit all the reflectivity indexes, Eq. (\ref{reflex}), tend to unity and hence in the zero-point energy for the ideal conducting waveguide is just given  by
\begin{align}
    E^{Ideal}_{ZP}=&\left(\frac{\hbar L_z}{4\pi^2}\right)\int_{-\infty}^{\infty}\!\!\!\!\!\!dk\sum_{n=1}^\infty
    \int_{-\infty}^\infty d\xi\bigg\{a{\lambda}^{(0)}+b{\upsilon}^{(0)}\nonumber\\
    &+\ln\left(1-e^{-2a{\lambda}^{(0)}}\right)+
\ln\left(1-e^{-2b{\upsilon}^{(0)}}\right)\bigg\}.
\label{zpideal}
\end{align}
The above expression is symmetric under the permutation of the cavity length, so we can write
\begin{equation}
    E^{ideal}_{ZP}=\left(\frac{\hbar c L_z}{4\pi^2}\right)\bigg({\cal I}(a,b)+{\cal I}(b,a)\bigg),
    \label{zpideal2}
\end{equation}
with the integral
\begin{align}
    {\cal I}(a,b)=
    \sum_{n=1}^\infty
    \int d^2\rho&\bigg\{
    \,a\sqrt{\left(\frac{n\pi}{b}\right)^2+\rho^2}\nonumber\\
    &+\ln\left(1-e^{-2a\sqrt{\left(\frac{n\pi}{b}\right)^2+\rho^2}}\right)\bigg\},
    \label{Iab}
\end{align}
%The double integration in $\xi$ and $k$ can be thought as an  integration on a bi-dimensional coordinate system. In order to do this, first let us define $\eta=\xi/c$ and the vector $\rho=(k,\eta)$ with norm $\rho^2=k^2+\xi^2/c^2$. In this manner we obtain explicitly the zero-point energy for the waveguide with perfect conducting surfaces
%\begin{widetext}
 % \begin{align}
  %  E^{0}_{ZP}=\left(\frac{\hbar c L_z}{2\pi^2}\right)\sum_{n=1}^\infty
   % \int d^2\rho\,\,
   % \bigg\{&a\sqrt{\left(\frac{n\pi}{b}\right)^2+\rho^2}+\frac{1}{2}\ln\left(1-e^{-2a\sqrt{\left(\frac{n\pi}{b}\right)^2+\rho^2}}\right)\nonumber\\ &
   %+b\sqrt{\left(\frac{n\pi}{a}\right)^2+\rho^2}+\frac{1}{2}\ln\left(1-e^{-2b\sqrt{\left(\frac{n\pi}{a}\right)^2+\rho^2}}\right)
   %\bigg\}.
   %\label{ZP ideal}
%\end{align}  
%\end{widetext}
where we have defined the bidimensional vector $\vec{\rho}=(k,\xi/c)$ with $\rho^2=k^2+\xi^2/c^2$.
The ideal result of the zero-point energy given by Eq. (\ref{zpideal}) need to be regularized.
%, since formally the expression Eq. (\ref{Iab}) gives an infinite result. In order to obtain a finite Casimir energy we need to realize a regularization process. 
Here we use the dimensional regularization \cite{dim2, Ashmore:1972uj, Edery:2005bx}. In this manner, let us define the $s$-dimensional integrals
\begin{align}
   {\cal J}_s(a,b)&= a\,\sum_{n=1}^\infty
    \int d^s\rho\,\,
    \sqrt{\left(\frac{n\pi}{b}\right)^2+\rho^2},\nonumber\\
    {\cal K}_s(a,b)&=\sum_{n=1}^\infty
    \int d^s\rho\,\,\ln\left(1-e^{-2a\sqrt{\left(\frac{n\pi}{b}\right)^2+\rho^2}}\right),
    \label{JsKs}
\end{align}
from these expressions, we see that the zero-point energy Eq. (\ref{zpideal}) can be recovered when $s=2$, since it is clear that
${\cal I}(a,b)={\cal J}_2(a,b)+{\cal K}_2(a,b)$.
The terms in Eq. (\ref{JsKs}) only depends on the modulus of the $s$-dimensional vector $\vec{\rho}$, one can perform the general solid angle integration with Eq. (\ref{solid angle}).
%$$\int d^s\rho=\frac{2\pi^{s/2}}{\Gamma(s/2)}\int_0^\infty %d\rho\,\rho^{s-1},$$
%where $\Gamma$ is the Euler gamma function. 
For other side, the integration over the modulus $\rho$ can be realized by using the Beta function  representations%properties
$$B(x,y)=\frac{\Gamma(x)\Gamma(y)}{\Gamma(x+y)}.$$
Finally, by using the reflection formula of the Riemann zeta function,  analogously to the Eq. (\ref{eq:rie}),
%$$\Gamma\left(\frac{s}{2}\right)\pi^{-s/2}\zeta(s)=\Gamma\left(\frac{1-s}{2}\right)\pi^{(s-1)/2}\zeta(1-s),$$
one can prove that
\begin{equation}
    {\cal J}_s(a,b)=-\frac{a}{2b^{s+1}}\pi^{s/2-1}\Gamma\left(\frac{s+2}{2}\right)\zeta(s+2).
\end{equation}

The second regularized integral ${\cal K}_s$, in Eq. (\ref{JsKs}), can be rewritten by using the  Taylor expansion of the logarithm function as follows
$\ln(1-x)=-\sum_{k=1}^\infty x^k/k$ valid for $|x|<1$. Hence, we can write
\begin{equation}
    {\cal K}_s(a,b)=\!-\frac{2\pi^{s/2}}{\Gamma(s/2)}\sum_{n=1}^\infty
    \sum_{m=1}^\infty\frac{1}{m}\!\!\int_0^\infty \!\!\!\!\!d\rho\,\rho^{s-1}e^{-2ma\sqrt{\left(\frac{n\pi}{b}\right)^2+\rho^2}}.
    \nonumber
\end{equation}
In order to perform the integration above, we use
 the integral representation of the modified Bessel function \cite{Gradshteyn:1943cpj}
$$K_\nu(xz)=\frac{\sqrt{\pi}}{\Gamma(\nu+\frac{1}{2})}\left(\frac{x}{2z}\right)^\nu\int_0^\infty dt\,t^{2\nu}\frac{e^{-x\sqrt{t^2+z^2}}}{\sqrt{t^2+z^2}},$$
hence, one obtain that
\begin{equation}
    {\cal K}_s(a,b)=\frac{2\pi^{\frac{3s-1}{2}}}{b^s}\sum_{n=1}^\infty
    \sum_{m=1}^\infty\frac{1}{m}
    \frac{d}{d\lambda}\left(K_{\frac{s-1}{2}}(\lambda n)\left(\frac{2n}{\lambda}\right)^{\frac{s-1}{2}}\right),
    \nonumber
\end{equation}
where we have defined $\lambda=2m\pi a/b$. Now, by using the following recursion formula of modified Bessel function
$$\left(\frac{1}{z}\frac{d}{dz}\right)^mK_0(z)=(-1)^mz^{-m}K_m(z),$$
and also considering the series expansion
\begin{align}
  \sum_{n=1}^\infty K_0(\lambda n)&=\frac{1}{2}\left(C+\ln\left(\frac{\lambda}{4\pi}\right)\right)+
\frac{\pi}{2\lambda}
\nonumber\\&
+\pi\sum_{p=1}^{\infty}
\left(\frac{1}{\sqrt{\lambda^2+(2\pi p)^2}}-\frac{1}{2\pi p}\right),  
\end{align}
one can obtain that
\begin{align}
    {\cal K}_s(a,b)=&2\pi^{\frac{s-1}{2}}\bigg\{
    \frac{1}{4a^s}\Gamma\left(\frac{s+1}{2}\right)\zeta(s+1)\nonumber\\
    &+\frac{1}{4\sqrt{\pi}}\Gamma\left(\frac{s+2}{2}\right)\zeta(s+2)\frac{a}{b^{s+1}}\nonumber\\
    &-\frac{1}{8\sqrt{\pi}}\Gamma\left(\frac{s+2}{2}\right)ab\,Z_2(a,b;s+2)
    \bigg\},
\end{align}
where we have used the definition of the Epstein zeta function, Eq. (\ref{eq:eps}). Now, putting all together, one can recast Eq. (\ref{zpideal2}) as 
%In the above equation we have introduce the Epstein zeta function
%$$Z(a,b,s)=\sum_{k=1}^\infty\sum_{p=1}^\infty\left((ak)^2+(bp)^2\right)^{-s/2}.$$
\begin{equation}
    E^{Ideal}_{C}=\frac{\hbar cL_z}{8\pi^2}\left(
    \frac{\pi}{2}\zeta(3)\left(\frac{1}{a^2}+\frac{1}{b^2}\right)-abZ_2(a,b,4)
    \right).
    \label{zpideal3}
\end{equation}
By using the variable $r=a/b$ one can express this energy as
$$E_C^{Ideal}=\left(\frac{\hbar cL_z}{a^2}\right){\cal E}(r),$$ with the adimensional function
\begin{equation}
    {\cal E}(r)=
    \frac{1}{16\pi}\zeta(3)(1+r^2)-\frac{r^3}{8\pi^2}Z_2(r,1,4).
\end{equation}
With the Casimir energy at hands, one can find the Casimir force 
\begin{equation}
    F_C^{Ideal}=-\frac{\partial E_C^{Ideal}}{\partial a}.
\end{equation}
We can write it in terms of adimensional variable $r$ as
\begin{equation}
    F_C^{Ideal}=\left(\frac{\hbar cL_z}{a^3}\right)f(r),
\end{equation}
where 
\begin{equation}
    f(r)=\frac{\zeta(3)}{8\pi}-\frac{\zeta(4)}{\pi^2 r}+\frac{r^3}{8\pi^2}Z_2(r,1,4)-\frac{2r^3}{\pi^2}z(r),
\end{equation}
and we have defined the function $z(r)$ given by
\begin{equation} z(r)=\sum_{m,p=1}^\infty\frac{(mr)^2}{\left[(mr)^2+p^2\right]^3}.
\end{equation}
\begin{centering}
    \begin{figure}[h]
        \centering
        \includegraphics[scale=0.55]{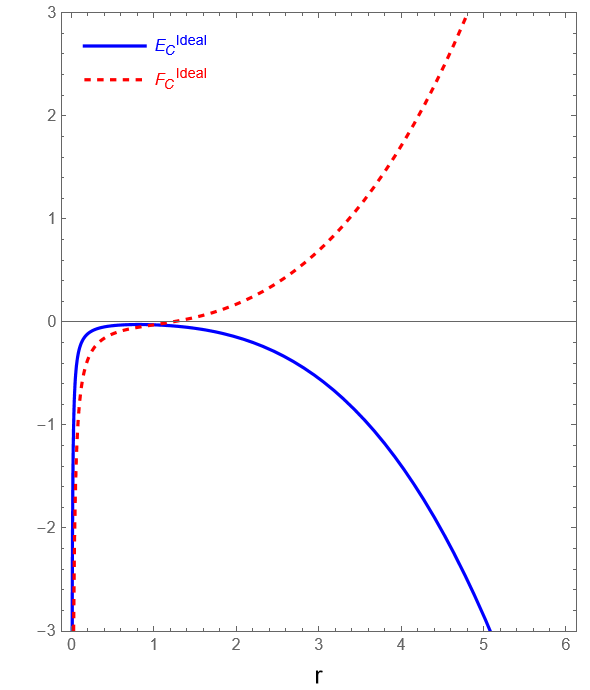}
        \caption{Casimir energy and force in the case of waveguide with perfect conducting surfaces. These quantities are plotted as a function of the ratio $r=a/b$ between the length of the waveguide cross-section. The energy is measure in units of $\hbar cL_z/a^2$ while force units are $\hbar cL_z/a^3$.} 
        \label{EC FC ideal}
    \end{figure}
\end{centering}

In Fig. (\ref{EC FC ideal}) we show the graph of the Casimir energy and the Casimir force for a perfectly conducting waveguide of rectangular cross-section. The Casimir energy is always negative but exhibits a maximum value for some critical value $r_{c}$ of the ratio between the waveguide lengths $r=a/b$. Near that point, the Casimir force is null and changes its behavior from attractive for $r<r_{c}$, to repulsive for $r>r_{c}$. The dependence of the attractive-repulsive nature of the Casimir force with the length ratio of the cavity shape is a known in the literature that we are recovering in the limit of perfect conductivity \cite{LUKOSZ1971109, Caruso:1998ru}.

%\subsection{Imperfectly conducting surfaces corrections}
%In this section we obtain 
We conclude this section, presenting the corrections to the zero-point energy due to a finite conductivity of the waveguide material. First, we define some useful variables. In this way, instead of working with the variables $\lambda_{1,2}$ and $\upsilon_{1,2}$, see Eq. (\ref{lambda}) and Eq. (\ref{upsilon}), we define the variables $p$ and $\tilde{p}$ given by
\begin{align}
    \left(\frac{n\pi}{b}\right)^2+k^2&=\epsilon_1\frac{\xi^2}{c^2}(p^2-1),\nonumber\\
     \left(\frac{n\pi}{a}\right)^2+k^2&=\epsilon_1\frac{\xi^2}{c^2}(\tilde{p}^2-1),
\end{align}
for the other side, we define the variables $s$ and $\tilde{s}$ as
\begin{align}
    p^2-1+\frac{\epsilon_2}{\epsilon_1}=s^2,\nonumber\\
    \tilde{p}^2-1+\frac{\epsilon_2}{\epsilon_1}=\tilde{s}^2,
\end{align}
with these variables one can show that $\lambda_1=\sqrt{\epsilon_1}\xi p/c$, $\lambda_2=\sqrt{\epsilon_1}\xi s/c$, $\upsilon_1=\sqrt{\epsilon_1}\xi \tilde{p}/c$, and $\upsilon_2=\sqrt{\epsilon_1}\xi \tilde{s}/c$. Hence, the reflectivity index, Eq. (\ref{reflex}), can be rewritten as
\begin{align}
    {\cal X}_1&=\left(\frac{p\epsilon_2+s\epsilon_1}{p\epsilon_2-s\epsilon_1}\right)^2,\hspace{.5cm}
    {\cal X}_2=\left(\frac{\tilde{p}+\tilde{s}}{\tilde{p}-\tilde{s}}\right)^2,\nonumber\\    
    {\cal Y}_1&=\left(\frac{\tilde{p}\epsilon_2+\tilde{s}\epsilon_1}{\tilde{p}\epsilon_2-\tilde{s}\epsilon_1}\right)^2,\hspace{.5cm}
    {\cal Y}_2=\left(\frac{{p}+{s}}{{p}-{s}}\right)^2.
    \label{reflex plasma}
\end{align}
We showed in the previous section that this reflectivity indexes tend to unity in the case of vacuum waveguide with perfect conductor in the outside ${\cal X}_{1,2}, {\cal Y}_{1,2}\rightarrow 1$. 
For the case of imperfect conducting surface we expect a behavior of the kind that give us some finite corrections to the ideal perfect conducting case, such as
\begin{equation}
    {\cal X}_{1,2}=1+\Delta {\cal X}_{1,2},\hspace{.7cm}  {\cal Y}_{1,2}=1+\Delta {\cal Y}_{1,2}.
\end{equation}
Considering small corrections, we can expand for the imperfect conductor case the zero-point energy as
\begin{equation}
    E_{ZP}=E^{Ideal}_{ZP}+\Delta E_{ZP},
\end{equation}
where the ideal case is given by Eq. (\ref{zpideal}) or equivalently by Eq. (\ref{zpideal3}). The correction due to finite conductivity of the dielectric is given by the term $\Delta E_{ZP}$. Explicitly, in this case we assume vacuum inside the waveguide with dielectric constant $\epsilon_1=1$, while for the outside we consider that the dielectric constant follows the plasma model and is given by $\epsilon_2=\epsilon(\omega)$, where
\begin{equation}
    \epsilon(\omega)=1-\frac{\omega_p^2}{\omega^2},
    \label{plasma model}
\end{equation}
and the plasma frequency $\omega_p$ is a specific characteristic of the material. Considering this, we have that in terms of the imaginary frequency  $\epsilon_2=1+\omega_p^2/\xi^2$.

\begin{centering}
    \begin{figure}[h!]
    \centering
    \includegraphics[scale=0.65]{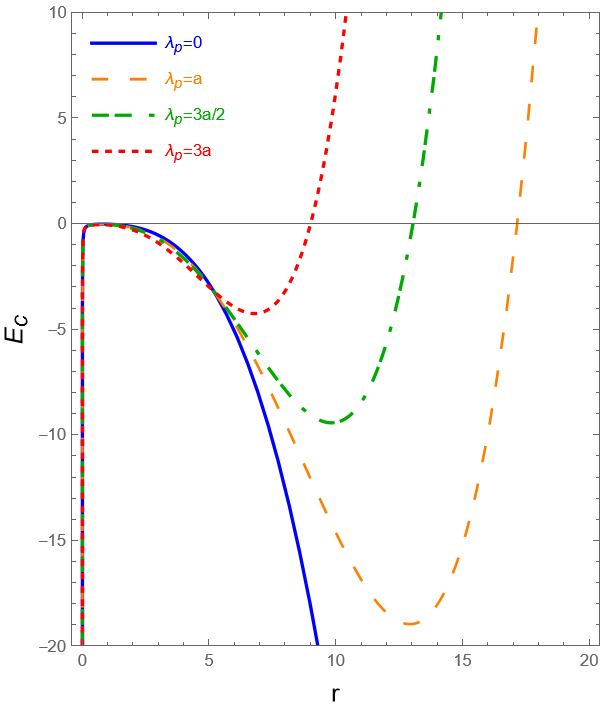}
    \caption{The final Casimir energy corrected by the finite dielectric properties of the material. The correction depend on the plasma wave length $\lambda_p$ with respect to $a$. The energies are in units of $\hbar c L_z/a^2$ and $r=a/b$ is the ratio between the length of the cross-section.}
    \label{corrected casimir energy}
\end{figure}
\end{centering}

\begin{centering}
    \begin{figure}[h!]
    \centering
    \includegraphics[scale=0.65]{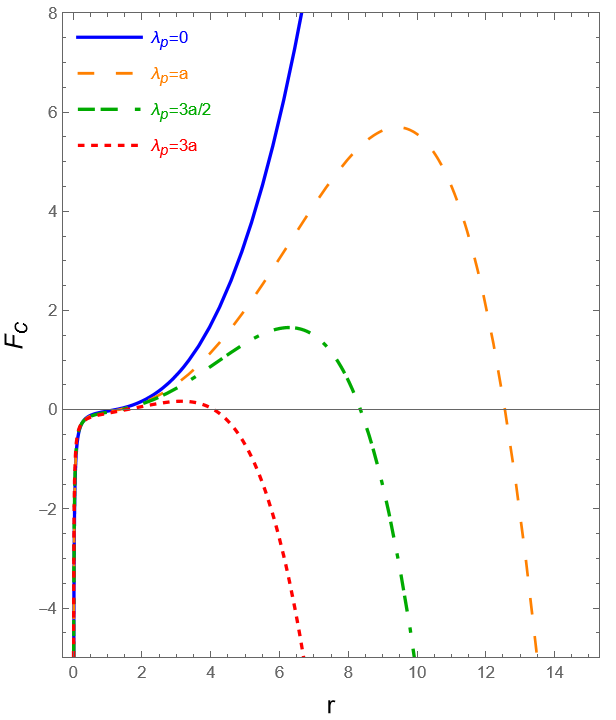}
    \caption{The final Casimir force corrected by the finite dielectric properties of the material. The correction depend on the plasma wave length $\lambda_p$ with respect to $a$. The forces are in units of $\hbar c L_z/a^3$ and $r=a/b$ is the ratio between the length of the cross-section.}
    \label{corrected casimir force}
\end{figure}
\end{centering}

Since inside the waveguide we have vacuum, then the wave number there is $\lambda_1\rightarrow\lambda^{(0)}$, see 
Eq, (\ref{lambda upsilon 0}), and one can define $p_{0}$ such that
$\lambda^{(0)}=\xi p_{0}/c$ and 
 $s=\sqrt{p_0^2+\omega_p^2/\xi^2}$, 
 by expanding $s$ up to first order in $\xi/\omega_p$, we obtain
\begin{equation}
    s\approx\frac{\omega_p}{\xi}+\frac{p_0^2\,\xi}{2\omega_p},
\end{equation}
by using the above expansion together with Eq. (\ref{plasma model}) and that $\epsilon_1\rightarrow1$ with $p\rightarrow p_0$ one find that the corrections to the reflectivity indexes
\begin{equation}
     \Delta {\cal X}_1\approx\frac{4\xi}{p_0\,\omega_p },\hspace{.6cm}
    \Delta {\cal Y}_2\approx\frac{4p_0\,\xi}{\omega_p}.
\end{equation}
In a very similar way, we have that inside the waveguide 
$\upsilon_1\rightarrow\upsilon^{(0)}$, and with this we define the variable $\tilde{p}_{0}$ such that
$\upsilon^{(0)}=\xi \tilde{p}_{0}/c$ and 
 $\tilde{s}=\sqrt{\tilde{p}_0^2+\omega_p^2/\xi^2}$. By realizing the expansion, one obtain the first corrections to the reflectivity indexes
\begin{equation}
     \Delta {\cal Y}_1\approx\frac{4\xi}{\tilde{p}_0\,\omega_p },\hspace{.6cm}
    \Delta {\cal X}_2\approx\frac{4\tilde{p}_0\,\xi}{\omega_p}.
\end{equation}
By considering this finite conductivity corrections, one can write the zero-point energy correction as given by
\begin{widetext}
\begin{equation}
    \Delta E_{ZP}=\left(\frac{\hbar L_z}{8\pi^2}\right)\int_{-\infty}^{\infty}dk\sum_{n=1}^\infty
    \int_{-\infty}^\infty d\xi\bigg\{
    (\Delta {\cal X}_1+\Delta {\cal Y}_2)\left(1-e^{-2a\lambda^{(0)}}\right)^{-1}
    +(\Delta {\cal X}_2+\Delta {\cal Y}_1)\left(1-e^{-2b\lambda^0}\right)^{-1}
    \bigg\}.
\end{equation}
By exploring the symmetry of this energy correction under the permutation of the cavity length $a$ and $b$, one can express this as a function of only the scale $a$ and on the ratio $r=a/b$. Also, considering that the plasma wavelength of the material is $\lambda_p=2\pi c/\omega_p$, we can write
\begin{equation}
    \Delta E_{ZP}=\hbar L_z c\lambda_p\left(\frac{{\cal U}(r)}{a^3}+\frac{{\cal U}(r^{-1})}{b^3}\right),
\end{equation}
where we have defined the adimensional function
\begin{equation}
{\cal U}(r)= \frac{1}{2\pi^2}
\sum_{n=1}^\infty
    \int_{0}^\infty d\chi\,\chi\,
    \left(\sqrt{(n\pi r)^2+\chi^2}+\frac{\chi^2}{2\sqrt{(n\pi r)^2+\chi^2}}\right)\left(\frac{1}{1-e^{-2\sqrt{(n\pi r)^2+\chi^2}}}\right),  
\end{equation} 
\end{widetext}
%we can use the symmetry of the result and write teh finite conductivity correction as 
%\begin{equation}
 %   \Delta E_{ZP}={\cal E}(a,b)+{\cal E}(b,a),
%\end{equation}
%where
%\newpage
%\begin{widetext}
%\begin{equation}
%{\cal E}(a,b)=  \left(\frac{2\hbar c L_z}{\pi^2\omega_p}\right)
%\int_{0}^{\infty}dk\sum_{n=1}^\infty
%    \int_{0}^\infty d\xi
%    \left(\gamma^0+\frac{\xi^2}{c^2 \gamma^0}\right)\left(\frac{e^{2a\gamma^0}}{e^{2a\gamma^0}-1}\right),  
%\end{equation}
%\end{widetext}
%by considering only term of first order in $\xi/\omega_p$, one can dismiss the second term inside the integral above.
%Also, by using the polar variables $\eta$ and $\rho$, as defined previously, we can write explicitly
%\begin{widetext}
%\begin{equation}
%{\cal E}(a,b)\approx \left(\frac{\hbar\, c^2 L_z}{\pi\omega_p}\right) 
%\sum_{n=1}^\infty
%    \int_{0}^\infty d\rho\,\rho\,
%    \sqrt{\left(\frac{n\pi}{b}\right)^2+\rho^2}\left(\frac{e^{2a\sqrt{\left(\frac{n\pi}{b}\right)^2+\rho^2}}}{e^{2a\sqrt{\left(\frac{n\pi}{b}\right)^2+\rho^2}}-1}\right). 
%\end{equation}
%\end{widetext}
%If we define the ratio between the lengths of the transverse section of the waveguide as $r=a/b$ and the adimensional variable $\chi=a\rho$ one can write the function as
%\begin{equation}
 %   {\cal E}(a,b)=\left(\frac{\hbar\, c^2 L_z}{\pi\omega_p\,a^3}\right){\cal U}(r),
%\end{equation}
as expected, this integral need to be regularized. We use dimensional regularization in the same line as realized before for the case of ideal perfect conducting waveguide. %After doing this we find that
We find that
\begin{equation}
    {\cal U}(r)=
    -\frac{3\zeta(3)}{32\pi^2}
    +\frac{\zeta(4)}{16\pi^3 r}+\frac{r^3}{2\pi^3}z(r),
\end{equation}
one can express 
\begin{equation}
    \Delta E_{ZP}=\left(\frac{\hbar L_z c\lambda_p}{a^3}\right)\Delta{\cal E}(r),
\end{equation}
where $\Delta{\cal E}(r)={\cal U}(r)+r^3{\cal U}(r^{-1})$. 
%With this we can obtain the final expression for the Casimir energy in the dielectric waveguide as
Which allow us to write the Casimir-Lifshitz energy for the dielectric wave guide in the plasma model as
\begin{equation}
    E_{C}=\left(\frac{\hbar cL_z}{a^2}\right)\left(
    {\cal E}(r)+\frac{\lambda_p}{a}\Delta{\cal E}(r)\right).
\end{equation}
The Casimir-Lifshitz energy is shown at Fig. (\ref{corrected casimir energy}) for different values of the plasma wavelength $\lambda_p$ measure in units of the length $a$.
In this figure, we note that the effect of the finite conductivity correction is to change the concavity of the Casimir energy for large values of $r$. This would lead to the appearance of a local minimum of the zero-point energy near some value $r^*$. Near this point 
the Casimir-Lifshitz energy behaves as an effective potential well.
%\st{and the vacuum fluctuations may generate natural vibrations on the waveguide walls, in the case of two non-fixed parallel walls. The generation of the natural vibration induced only by vacuum energy is more evident when we analyze} 
In this case
the Casimir force given by Fig. (\ref{corrected casimir force}).
In that figure, a second inversion on the attractive-repulsive nature of the Casimir force occurs at $r^*$. This second point is also an equilibrium point with zero Casimir force, but it corresponds to a stable equilibrium, whereas the initial critical point, $r_{c}$, is an unstable equilibrium. %\st{This behavior brings a possibility to construct a Casimir-Lifshitz anharmonic oscillator.}

\section{Conclusions}\label{sec:conclusion}

%\st{The literature has been discussed MEMS devices using the Casimir effect. A particular interesting device was proposed by Serry et. al }
%\cite{ACOmaclay1995}. 
%\st{They proposed an anharmonic Casimir oscillator constructed with two parallel plates, where one of the plates is connected to a surface with a spring.} 
%\st{Here we show that it is possible to obtain a Casimir-Lifshitz anharmonic oscillator where the restoring force is driven only by vacuum fluctuation effects.}

In this paper we present an analysis of the Casimir-Lifshitz effect associated with the electromagnetic field confined within a rectangular waveguide consisting of two distinct dielectric materials. 
We employ the surface mode technique to derive a generalized Lifshitz formula for this specific geometry. Our formulation accounts for the unique dielectric properties of the materials composing the waveguide, leading to a precise calculation of the Casimir-Lifshitz energy. 
%{\color{red}The behavior of the force on the walls on the rectangular cavity allow us to identify the system as a Casimir-Lifshitz anharmonic oscillator.}
%, where, in the case of two non-fixed walls, these two walls}
%\st{The zero-point energy has a local minimum. Therefore, oscillatory systems may be constructed.}
The behavior of the force on the walls of the rectangular cavity  open the possibility to construct  a Casimir-Lifshitz anharmonic oscillator.
%One can discuss further the case where rigid plates are allowed to move. We are interpreting this as the Casimir-Lifshitz anharmonic oscillator.}vibrate due to only the vacuum stress.
A natural continuation of this work is to generalized our calculations considering a dissipative model for the dielectric dependence on the wave frequency \cite{barton2001perturbative}.

\appendix
\section{boundary conditions}\label{app:A}
In this appendix, we analyze the correct boundary conditions to be satisfied for the electric and magnetic fields in the rectangular dielectric waveguide. We assume that there is no free charge or current in the materials interfaces. Therefore, in a given boundary surface between the materials, the normal component of the vector fields ${\bf D}=\epsilon{\bf E}$ and ${\bf B}$ must be continuous, as well as the tangential components of ${\bf E}$ and ${\bf B}$. It is clear that the correct direction of normal and tangential components depend on the specific surface we are referring to. In the following lines, we indicate  explicitly the boundary conditions of the vector fields for each surface that compose the waveguide.\\

\noindent
\textit{\bf Boundary conditions for verticals $V_1$ and $V_2$}:\\
\noindent
The normal component of the electric displacement vector ${\bf D}$ must be continuous across these surfaces. Therefore, this implies the continuity of $\epsilon(\omega)\,e_x$ at the vertical surfaces  $V_1$ and $V_2$. The continuity of the tangent components of the electric field imply that $e_y$ and $e_z$ must be also continuous. The continuity of the $z$-component together with Eq. (\ref{divE}) implies that $\partial_xe_x+\partial_ye_y$ must be continuous. This final result could be satisfied if, independently, we have that $\partial_x\,e_x$ and $\partial_y\,e_y$ are continuous also.

The continuity of the normal component of the magnetic field indicates that on the vertical surfaces the term  $(\partial_y\,e_z-ike_y)$ must be continuous, which would add the continuity condition of $\partial_y\,e_z$.
The continuity of the tangent component of the magnetic field in the $z$-direction implies that $(\partial_y\,e_x-\partial_x\,e_y)$ is continuous. We can satisfy this conditions if $\partial_y\,e_x$ and $\partial_x\,e_y$ are continuous. The continuity of the tangent component of the magnetic field in the $y$-direction indicates that $(ik\,e_x-\partial_xe_z)$ must be continuous. But we can rewrite this term by using Eq. (\ref{divE}), as follows
\begin{align}
    ik\,e_x-\partial_xe_z&=ik\,e_x+\frac{1}{ik}\partial_x(\partial_xe_x+\partial_ye_y)\nonumber\\
    &=\frac{1}{ik}(-k^2e_x+\partial^2_xe_x+\partial^2_{x,y}e_x)\nonumber\\
    &=\frac{1}{ik}\left(-\frac{\omega^2}{c^2}\epsilon(\omega)e_x+\partial_y(\partial_xe_y-\partial_ye_x)\right),
\end{align}
where we have used the wave equation Eq. (\ref{we}). This last conditions is satisfied since $\epsilon(\omega)\,e_x$ is already continuous.\\

\noindent
\textit{\bf Boundary conditions for horizontals $H_1$ and $H_2$}:\\
\noindent
The continuity of the normal component of ${\bf D}$ implies the continuity of $\epsilon(\omega)\,e_y$ at the horizontal surfaces of the waveguide. The continuity of the tangent components of the electric field imply that $e_x$ and $e_z$ must be also continuous. The continuity of the $z$-component together with Eq. (\ref{divE}) implies that $\partial_xe_x+\partial_ye_y$ must also be continuous. This result could be satisfied if independently we have that $\partial_x\,e_x$ and $\partial_y\,e_y$ are continuous on the frontier.

For the magnetic field, the continuity of the normal component at the horizontal surfaces implies the continuity of $(ike_x-\partial_x\,e_z)$, which would be satisfied only if $\partial_x\,e_z$. The continuity of the tangent component of the magnetic field in the $z$-direction implies that $(\partial_y\,e_x-\partial_x\,e_y)$ is continuous. We can satisfy this conditions if independently $\partial_y\,e_x$ and $\partial_x\,e_y$ are continuous. The continuity of the  tangent $x$-component of the magnetic field on the horizontals surfaces, indicates that $(\partial_ye_z-ike_y)$ must be continuous. But we can rewrite this term using Eq. (\ref{divE}), as
\begin{align}
    \partial_ye_z-ike_y&=
    -\frac{1}{ik}\partial_y(\partial_xe_x+\partial_ye_y)
    -ike_y\nonumber\\
    &=
    -\frac{1}{ik}(\partial^2_ye_y-k^2e_y+\partial^2_{x,y}e_x)\nonumber\\
    &=\frac{1}{ik}\left(-\frac{\omega^2}{c^2}\epsilon(\omega)e_y+\partial_x(\partial_ye_x-\partial_xe_y)\right),
\end{align}
where we have used the wave equation Eq. (\ref{we}). This last conditions is satisfied since $\epsilon(\omega)\,e_y$ is already continuous on these horizontal surfaces.

\begin{acknowledgements} 
  We would like to thanks B. F. Svaiter for the useful discussions. This work was partially supported by Conselho Nacional de Desenvolvimento Cient\'{\i}fico e Tecnol\'{o}gico - CNPq, the grant - 305000/2023-3 (N.F.S). G. O. H. thanks  Fundação Carlos Chagas Filho de Amparo à Pesquisa do Estado do Rio de Janeiro (FAPERJ) due the financial support.
\end{acknowledgements}

 \bibliography{main.bib}

\end{document}